# Topological Darkness: How to Design a Metamaterial for Optical Biosensing with Virtually Unlimited Sensitivity

G. Tselikov[1], A. Danilov[1], V. O. Shipunova[2], S. M. Deyev[2], A. V. Kabashin[1], A. N. Grigorenko[3]

[1] *Aix Marseille University, CNRS, UMR 7341 CNRS, LP3, Campus de Luminy – case 917, 13288, Marseille Cedex 9, France*

[2] *Shemyakin–Ovchinnikov Institute of Bioorganic Chemistry, Russian Academy of Sciences, 16/10 Miklukho-Maklaya St, Moscow 117997, Russia*

[3] *Department of Physics and Astronomy, University of Manchester, Manchester, M13 9PL, UK*

## Abstract

Due to the absence of labels and fast analyses, optical biosensors promise major advances in biomedical diagnostics, security, environmental and food safety applications. However, sensitivity of the most advanced plasmonic biosensor implementations has a fundamental limitation caused by losses in the system and/or geometry of biochips. Here, we report a "scissor effect" in topologically dark metamaterials which is capable of providing virtually unlimited *bona fide* sensitivity to biosensing thus solving the bottleneck sensitivity limitation problem. We explain how the "scissor effect" can be realized via a proper design of topologically dark metamaterials and describe strategies for their fabrication. To validate the applicability of this effect in biosensing, we demonstrate the detection of folic acid (vitamin important for human health) in the wide 3-log linear dynamic range with the limit of detection of 0.125 nM, which is orders of magnitude better than previously reported for all optical counterparts. Our work opens possibilities for designing and realising plasmonic, semiconductor and dielectric metamaterials with ultra-sensitivity to binding events.

## Introduction

The analysis of affinity binding interactions between a target analyte (e.g., antigen, protein, peptide, DNA, RNA segments) from a biological sample solution and its selective receptor (e.g., antibody, protein, peptide, etc.) immobilized on the surface presents one of key tasks in biomedical diagnostics (e.g., the detection of biomarkers of infections and cancers, cardio control, immune status, rational drug design), environmental and food safety (e.g., the monitoring of toxins or pathogens), and security applications[1]. Conventional label-based biosensing, currently used in hospitals and laboratories, implies the use of fluorescence or radio labels to mark analytes and thus report a biomolecular binding, but this approach is insufficiently precise due to the presence of reaction-interfering labelling step, costly in terms of required laboratory installations, and excessively long. An alternative is offered by optical transduction biosensing to record biomolecular interactions via the monitoring of optical refractive index (RI) associated with the increase of biofilm thickness, which enables one to immediately report the result of binding and obtain kinetic constants within minutes[2]. A paramount importance of such label-free approach was greatly magnified by recent pandemic, which revealed a critical lack of reliable, easy-to-use and mass-scale biochips that could give immediate accurate testing results. Profiting from medium-dependent optical excitation of free electron oscillations (plasmons) and a much enhanced plasmon-mediated electric field probing RI variations, plasmonic sensors form the core of label-free biosensing technology applied for the detection of a variety of critically important analytes[2-4]. However, currently available plasmonic biosensing architectures based on spectral (or angular) interrogation under Surface Plasmon Resonance (SPR) analytes[2,3] or Localized Plasmon Resonance (LPR)[4] have a major sensitivity bottleneck. Indeed, sensitivity of SPR and LPR sensing schemes in terms of spectral shift per bulk refractive index unit (RIU) change is of the order of $(3-10) \cdot 10^3$ nm/RIU[3] and $(2-5) \cdot 10^2$ nm/RIU[4], respectively, which conditions orders of magnitude inferior limit of detection (LOD) compared to label-based sensors. Such a bottleneck is related to a series of fundamental limiting factors, including high losses in plasmonic metals[5], low quality of resonances in the case of uncoupled LPRs[4,6], and structure geometry limitations in the case of advanced surface lattice resonances (SLR) over periodic nanoparticle arrays[7,8]. The sensitivity handicap of plasmonic biosensors can be fully compensated by using phase as a sensing parameter instead of spectral interrogation due to the presence of a sharp jump in the very minimum of resonant curve[9,10], but the implementation of ultrasensitive phase interrogation schemes requires properly designed plasmonic architectures and more complicated instrumental readouts.

We recently described a new phenomenon referred to as topological darkness (TD)[11,12], which consists in exactly zero light reflection/transmission from a dedicated optical system and is typically observed as a well-defined feature in the measured spectrum consisting in a drop in reflection/transmission with a point of zero light intensity. The absence of reflection/transmission is topologically protected under TD by spectral properties of optical constants of materials and the positioning of the zero reflection/transmission surface, and survives any imperfection in the fabrication of topologically dark structures (in the absence of diffuse scattered light). We then described a set of metamaterials (heterostructures[13] and nanostructures[11,14]) that possess TD and demonstrated that the generation of TD can result in extreme singularities of phase of light, which can be used in *phase* interrogation schemes to radically improve the sensitivity of plasmonic label-free biosensors[11,15,16].

Here, we further explore optical phenomena associated with the generation of TD in designed metamaterials focussing on the spectral response of TD feature to refractive index variations. We report a "scissor effect" and theoretically show that this effect could yield virtually unlimited spectral sensitivity in the recording of binding biosensing events. We explain how to design and fabricate topologically dark metamaterials (TDMs) for the implementation of the "scissor effect". To justify the applicability of such metamaterials in biosensing, we performed a label-free quantitative detection of vitamin folic acid (molecular weight 441.4 Da) as an example of a very small molecule important for human health. We show that TDMs were capable of monitoring folic acid in a wide range of concentrations (5 nM – 5000 nM), while the observed limit of detection (LOD) for TDM of 0.125nM was orders of magnitude better than that of previously used label-free and label-based methods. Our work opens possibilities for development fast, inexpensive and ultrasensitive label-free optical biosensors which could be based on plasmonic, semiconductor or dielectric materials and structures.

## Results

**Topological darkness in attenuated reflection geometry**

We start by recalling the main features of the phenomenon of topological darkness, which guarantees zero light intensity of light reflection (or transmission) at some angle of incidence and some wavelength for a dedicated structure. Here, we concentrate on TD in attenuated reflection geometry (ATR), which is the best suited for biosensing applications. Let us consider light reflection from a structure shown in Fig. 1a, where a thin TDM layer (which can be flat, nanostructured or heterostructured) is placed at the bottom of a prism with refractive index $n_1$ and is in contact with a studied medium of refractive index $n_3$ (which can be water or phosphate buffered saline buffer often used in biosensing). For a given light polarization, thickness of the TDM layer, an angle of incidence, $\theta$ and a light wavelength, $\lambda$, light reflection from the structure shown in Fig. 1a can be made exactly zero by adjusting values of the optical constants ($n(\lambda)$, $k(\lambda)$) of the TDM layer due to the nature of Fresnel coefficients for the whole structure[11]. Here, $n(\lambda)+ik(\lambda)=\sqrt{\varepsilon(\lambda)}$ is the complex refractive index of the TDM layer and $\varepsilon(\lambda)$ is the optical permittivity of the layer.

Hence, for fixed TDM thickness and an angle of incidence, we will obtain a curve of exactly zero reflection for the studied structure in the 3D space of ($n(\lambda)$, $k(\lambda)$, $\lambda$). If we allow the angle of incidence $\theta$ to change, we will obtain a zero reflection surface shown in Fig. 1b for a hypothetical material (the cyan surface). If the spectral curve of the optical constants for the TDM layer ($n(\lambda)$, $k(\lambda)$) shown in Fig. 1b by the red curve has two points from each sides of the zero reflection surface (see the points A and B in Fig. 1b), then, we will always have an intersection of the spectral curve of TDM optical constants with the zero reflection surface, which would guarantee a zero reflection point C (Fig. 1b) due to topology and the Jordan theorem[11]. The zero reflection point C is topologically protected in a sense that small imperfections in TDM fabrication will not change the relative positions of points A and B with respect to the zero reflection surface and, hence, the zero reflection point C will be still observed due to Jordan theorem[11] albeit at a slightly different wavelength and the angle of incidence. We have to stress that these considerations are only valid when diffuse scattering is small and one can apply Fresnel theory to the studied structure. The same considerations can also be applied for light transmission and guarantee topologically protected zero transmission points. Here we will concentrate on TD under ATR reflection due to effectiveness of this geometry in biosensing applications.

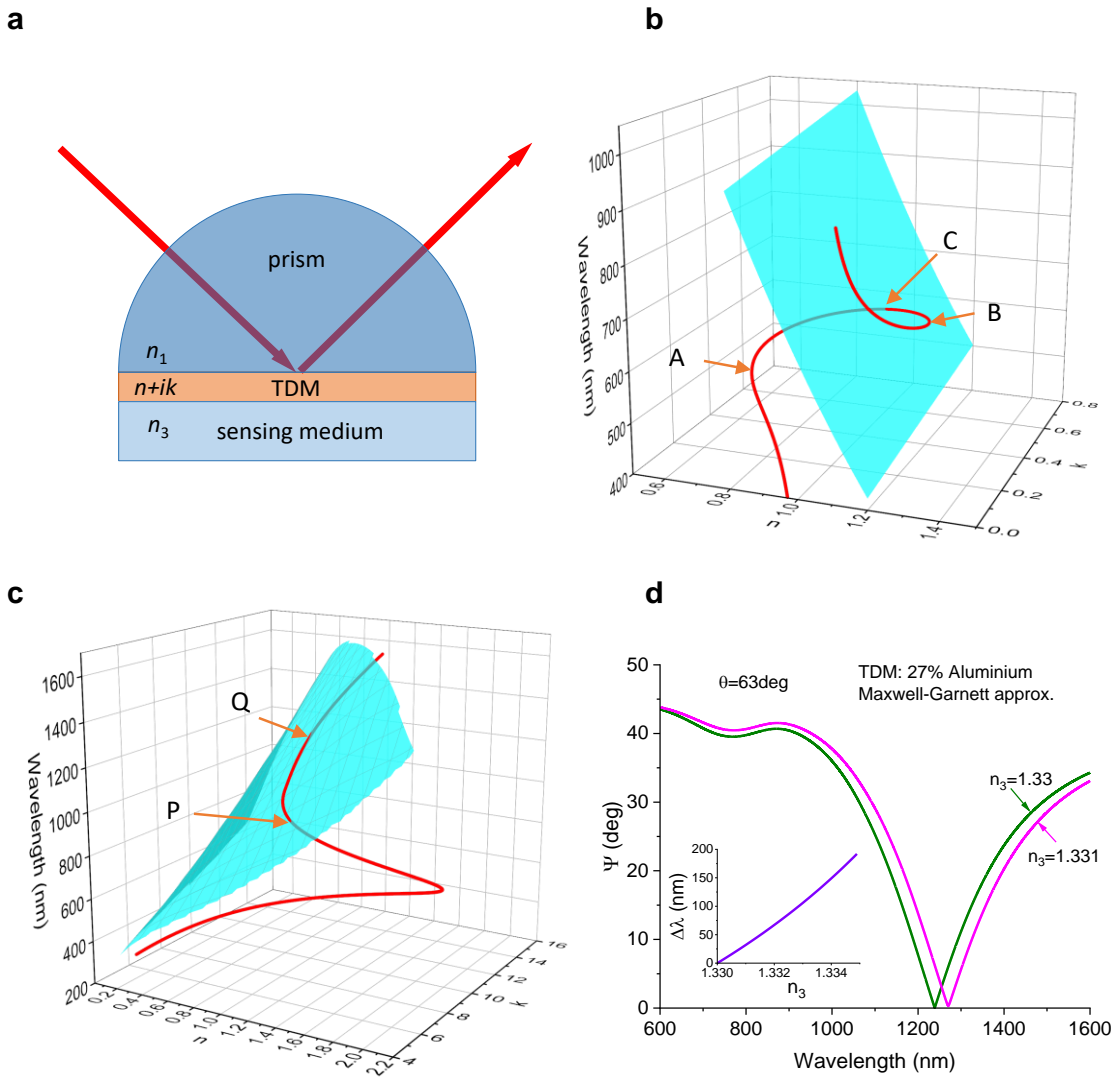

**Figure 1: Topological darkness in ATR geometry. a** The studied structure that consists of a coupling prism, a TDM layer and a sensing medium (e.g., water of PBS buffer). **b** Topologically protected darkness in the studied structure which is observed at the point C where the spectral optical constant curve of TDM ($n(\lambda)$, $k(\lambda)$), the red colour, intersects the zero reflection surface, the cyan colour. **c** The zero reflection surface (cyan colour) calculated for a structure of **a** for TDM of thickness of 21.696nm, $n_1$=1.513 and $n_3$=1.33. The red curve shows the spectral optical constant curve calculated for a TDM layer made of 27% Al and 73% of the bottom medium in the Maxwell-Garnett approximation. The red curve intersects the zero reflection surface at a small angle for the point Q. **d** The change of the spectral ellipsometric reflection $\Psi$ calculated for two different refractive indices of probed medium $n_3$=1.33 and $n_3$=1.331 (see Methods for the definition of $\Psi$) at a fixed angle of incidence θ=63.005° for the point Q of **c**. The insets show the spectral (amplitude) sensitivity of the TD as a function of the refractive index of the bottom medium. The thickness of TDM layer is 21.696nm.

**Sensitivity of TD materials to biosensing and the scissor effect**

The zero reflection point C presents a well-defined feature in the reflection spectrum and can be used for a label-free detection of biological binding events. A functionalization procedure of a TDM layer will provide selective detection of biological molecules resulting in the change of the effective thickness of TDMs (areal mass sensitivity) or in the change of the refractive index $n_3$ of the medium with which TDM is in contact (bulk sensitivity). In both cases, the zero reflection surface will shift and a new point of zero reflection will appear. The sensitivity of TDM amplitude detection then can be expressed in terms of $S = \Delta\lambda / \Delta n_3$, where $\Delta\lambda$ is the spectral shift of the zero reflection point caused by the change of the refractive index of the sensing medium $\Delta n_3$. It appears that this sensitivity should not be large as the shift of the zero reflection surface is normally small. However, in contrast to biosensing based on LPR or SPR platforms, where one measures a shift of a resonance curve with respect to its original position (which happens due to biological binding events in the probed medium or at the chip surface), *in the case of TD we measure how the zero reflection surface is moving with respect to the spectral curve (n(λ), k(λ)) of TDMs*. Hence, the sensitivity of TD structures crucially depends on the angle at which the spectral curve of TDM intersects the zero reflection surface. The smaller the angle of intersection, $\alpha$, the larger the spectral shift and the spectral sensitivity can be. We will refer to the large increase of sensitivity due to a small intersection angle as the "scissor effect" following a scissor analogy, where the point of intersection of scissor blades moves much faster than the blades themselves with an increase of the speed being proportional to $1/\alpha$ (where $\alpha$ is expressed in radians).

It is easy to find a metamaterial for which a spectral curve of optical constants (*n(λ)*, *k(λ)*) intersects the zero reflection surface at a small angle. Figure 1c shows a theoretical example of such situation for a structure depicted in Fig. 1a. In this case, a TDM of thickness *d*=21.696nm is produced by a mixture of Aluminium (27%) and the bottom medium (73%) described by Maxwell-Garnett theory[17]

(a particular type of effective medium is not important as this effect will be observed for, e.g., Bruggeman's effective medium[17] and others). This TDM has spectral optical parameters ($n(\lambda)$, $k(\lambda)$) shown in Fig. 1c by the red curve. The zero reflection surface for this TDM structure is shown in Fig. 1c by the cyan surface. We see that the spectral curve indeed intersects the zero reflection surface at a small angle for the point Q of zero reflection (it happens at the angle of incidence $\theta=63.005°$ and wavelength 1238.8nm). By calculating the shift of the zero reflection surface induced by a small change of the index of refraction of studied medium, we can find a new point Q' of intersection of the spectral curve with new zero reflection surface which gives us a new spectral position of zero reflection and a slightly different angle of incidence at which TD happens. This calculations yields high spectral sensitivity $S \approx 6 \cdot 10^3$ nm/RIU (where RIU is the refractive index unit) for the point Q of the discussed structure for the case where the angle of incidence is allowed to change in order to restore the TD.

It should be noted, however, that the angle of light incidence is often fixed in optical interrogation schemes. For this case, the sensitivity of the TD metamaterials can be even larger. For example, the Maxwell-Garnett TDM layer consisting of 27% of Aluminium and 73% of the bottom medium shows even higher spectral sensitivity of $S \approx 3.2 \cdot 10^4$ nm/RIU which can be deduced from Fig. 1d and its inset that depicts the shift of the spectral position of the reflection minimum caused by the changes of the refractive index $n_3$ of the sensing medium (a definition of ellipsomertic reflection used in the Fig. 1d is provided in Methods section) at a fixed angle of incidence $\theta=63.005°$. How this higher spectral sensitivity (which is comparable with SPR sensitivity of gold chips at this spectral range[3]) is connected to "spoof" SPRs will be discussed in future publications.

**Design of ultrasensitive topologically dark metamaterials**

The scenario described in Fig. 1c gives a simple theoretical algorithm for *designing ultrasensitive TDM (UTDM) with virtually unlimited spectral sensitivity to sensing medium* for a structure of Fig. 1a operating at a fixed angle. This algorithm is illustrated in Fig. 2. Step 1. Fix thickness of the TDM layer and the angle of incidence. (One can select TDM thickness at 5-30nm to avoid the reduction of the number of TD points, and the angle of incidence $\theta > \arcsin(n_3 / n_1)$ in order to realise ATR, in which only evanescent waves are present in the sensing medium.) For example, for Fig. 2 we selected the TDM thickness d=21.696nm and the angle of incidence $\theta$=63°. Step 2. Construct zero reflection curves at these parameters for two different refractive indices $n_3$ (these curves are shown in Fig. 2a for $n_3'$ =1.33 and $n_3''$ =1.331 by the cyan and grey colours, respectively). Step 3. Choose two points on these two zero reflection curves separated by large enough spectral distance $\Delta\lambda$, e.g., points M and N of Fig. 2a. Step 4. Design a metamaterial for which the optical constant spectral curve passes through the points M and N as shown in Fig. 2a by the red curve. By design, this metamaterial will show exactly zero reflection for both refractive indices of the sensing medium separated by the chosen spectral distance, see Fig. 2b. Hence, the constructed TDM will have spectral sensitivity $S = \Delta\lambda / \Delta n_3$ which can be, in principle, as large as required, in accordance with the "scissor effect". Figure 2b plots two reflection curves for the structure of Fig. 1a with UTDM spectrally described by the red curve of Fig. 2a. We observe a large shift of the zero reflection wavelength at a small change of the refractive index of sensing medium which results in TD spectral sensitivity of $S$=10$^5$ nm/RIU.

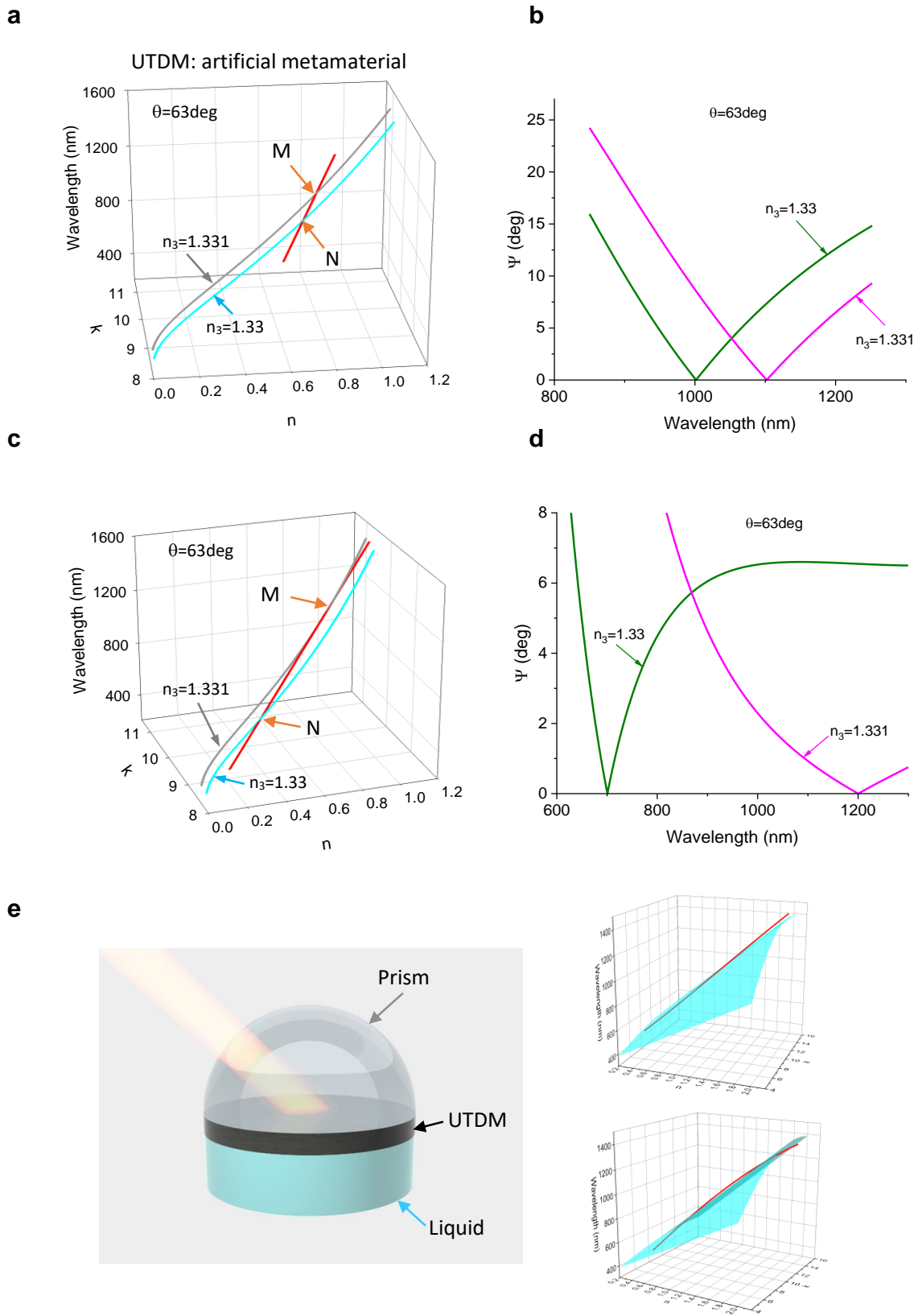

**Figure 2: Unlimited spectral sensitivity of topological dark metamaterials in biosensing. a** Zero reflection curves for two different refractive indices $n_3$ of sensing medium shown by the cyan and grey colour observed for the angle of incidence 63° and the TDM thickness of 21.696nm. The red curve show a spectral constant curve of a hypothetical metamaterial that connects points M and N. **b** The ellipsometric reflection Ψ calculated for two different refractive indices of probed medium $n_3$=1.33 and $n_3$=1.331 for TDM described by the red curve of **a** which shows a shift of the zero reflection point by 100nm. **c** Same as for **a** with larger spectral distance between points M and N. **d** Same as for **b** with a larger shift of the zero reflection point by 500nm. **e** The optics and geometry of UTDM. The top and bottom insets on the right show typical scenarios at which UTDM is observed where the cyan surface represents the zero reflection surface and the red curve represents the spectral curve of the optical constants of UTDM.

To demonstrate that UTDM sensitivity can be of any given number, Fig. 2c shows an application of the same recipe with a much larger spectral separation of points of M and N. Again, we design a metamaterial with the spectral curve of optical constants that pass through the points M and N shown as the red curve of Fig. 2c. This metamaterial yields even larger wavelength shift of 500nm for the zero reflection point at the change of the refractive index of sensing medium $\Delta n_3$ =0.001 resulting in sensitivity of $S=5\cdot10^5$ nm/RIU, see Fig. 2d. This sensitivity is about two orders of magnitude higher than that observed for optimised gold chips under SPR[3]. It clear that the suggested algorithm could provide "theoretical" metamaterial with basically any given number of spectral sensitivity.

## Discussion and conclusion

### Strategies to achieve the scissor effect

Figure 2e summarizes the main features of UTDM biosensing in the case of a simple structure shown in Fig. 1a, which comprises a coupling prism, a thin functionalized TDM layer, and sensing medium. The light falls on the structure at the angle of incidence which is larger than the critical angle of the structure, $\theta_c = \arcsin(n_3/n_1)$, producing only evanescent fields in the sensing medium (light transmission through the sensing medium is absent). The reflection from the structure is also absent due to TD. Hence, TDM behaves as a perfect absorber in this case[18,19]. To achieve the "scissor effect" and high bio-sensitivity, the spectral curve of UTDM should either intersect the zero reflection surface at a very small angle (as shown in the top inset of Fig. 2e) or almost touch it by crossing in two close points (see the bottom inset of Fig. 2e) leading to large spectral changes of the TD point at small changes of the sensing medium and opening limitless possibilities to design different types of UTDM. Finally, the small angle of intersection between the spectral curve and the zero reflection surface implies a small window of angles at which UTDM works which explains why this effect was not discussed before.

An important stage of designing UTDM is based on possibility to fabricate a metamaterial with given spectral parameters ($n(\lambda)$, $k(\lambda)$). This was a subject of intense studies in recent years driven by advances in nano-optics and plasmonics[20-25]. There are several approaches to design tailor-made metamaterials with a given spectral curve based on natural resonances[11], complex nanostructuring[11], SPR heterostructures[13], etc. These metamaterials can be based on plasmonic, semiconductor, or dielectric nano-heterostructures.

It is quite fortunate and counterintuitive that TDM that shows the "scissor effect" can be also realised even with reasonably simple nanostructured metal-dielectric metamaterials, e.g., formed by a regular square array of metal nanoparticles. Figure 3a shows SEM images of a simple TDM produced by a square array of gold dumbbells fabricated on a glass substrate (see Methods) which was used as a TDM layer in the structure of Fig. 1a. The measured ATR ellipsometric reflection of the structure with this TDM layer is shown in Fig. 3b (see Supplementary Information for the experimental details). This reflection has two main features – extremely narrow diffraction-coupled surface lattice resonance (SLR) at wavelength of 930 nm (where *s*-polarized reflection goes to zero) – and the point of topological darkness was observed at the wavelengths of 1185 nm and angle of incidence of 71°. The point of TD was observed in a narrow range of angle of incidences (the change of angle of incidence by 1° was enough to restore large reflection) which suggests that it can lead to the scenario described in Fig. 2e. Figure 3c shows that the curve of spectral constants of TDM extracted from Fig. 1b indeed intersects the zero reflection line at acute angle.

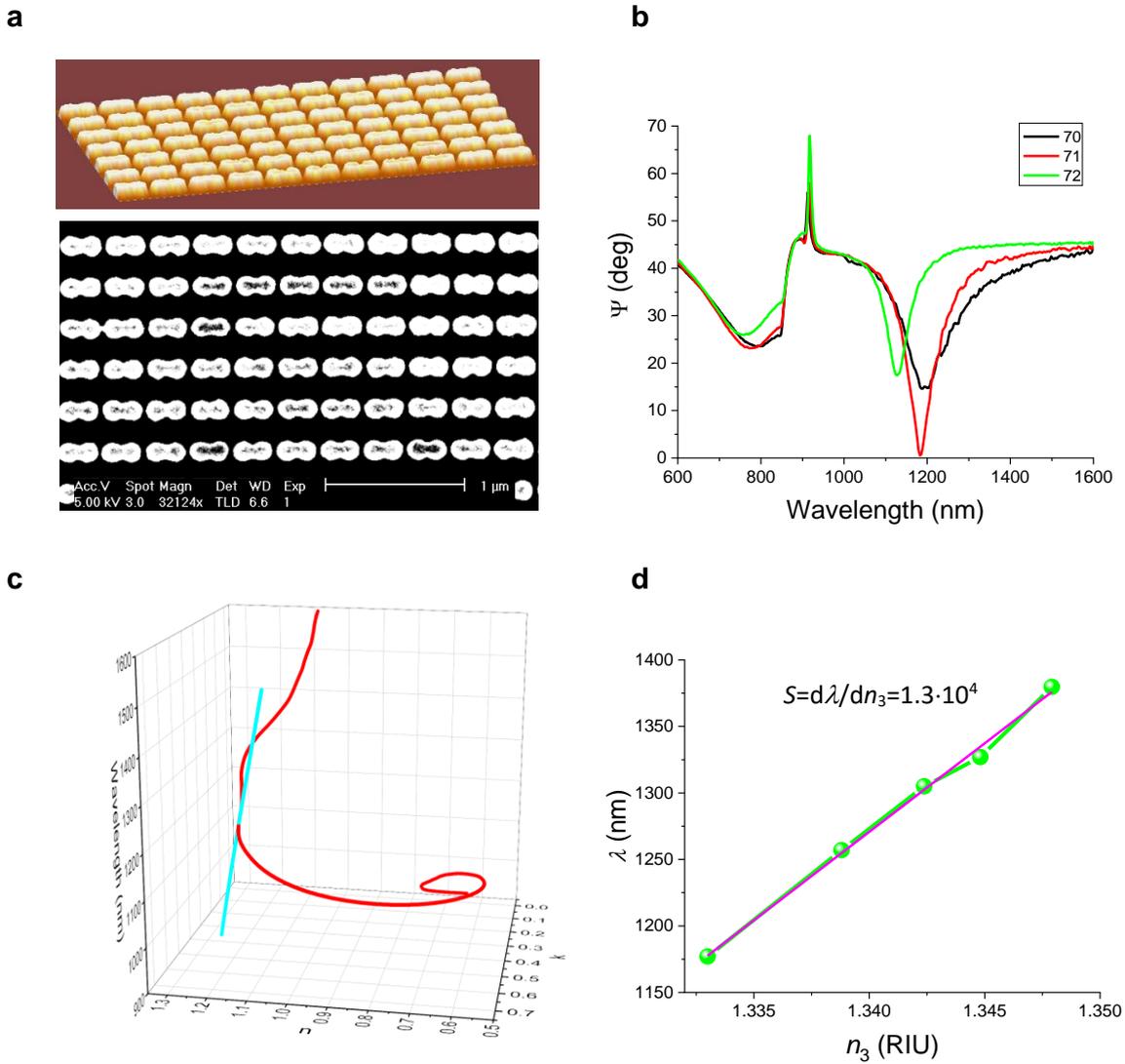

**Figure 3: The "scissor effect" and related enhanced spectral sensitivity of a fabricated topologically dark metamaterial based on a regular square array of Au nanoparticles. a** SEM images of the fabricated metamaterial. Top is a 3D image, bottom is a 2D image. **b** The ellipsometric reflection $\Psi$ measured for three angles of incidence for TDM shown in **a**. **c** An intersection of a zero reflection curve calculated at measured parameters (the cyan curve) with the spectral curve of optical constants of TDM extracted from **b**. **d** The change of the reflection minimum for the TDM structure of **a** measured using the water-glycerol mixtures (Methods).

To check the spectral sensitivity of TDM shown in Fig. 3a, we measured the experimental change of the reflection minimum by changing of the refractive index of sensing medium $n_3$. This was done using the water-glycerol mixtures (Supplementary Information) and resulted in experimental sensitivity of $S = 1.3 \cdot 10^4$ nm/RIU, which is more than two times higher than SPR sensitivity at these wavelengths, Fig. 3d. It is also worth comparing the sensitivity of the TD mode of the studied TD metamaterial with the sensitivity of SLR modes[15]. According to the theory, diffraction-coupled SLRs should provide sensitivity to RI variations at the level of $\Delta\lambda/\Delta n \sim d$, where $d$ is the array period[15]. This yields 320 nm/RIU for the SLR resonance (observed at 930 nm in the studied dumbbell array) which is 40 times smaller than the sensitivity of the TD mode, confirming its different nature.

**Ultrasensitive detection of folic acid with the help of topologically dark metamaterials**

To illustrate the applicability of TDMs in biosensing and the experimental usage of the "scissor effect" in biosensing, we carried out label-free biosensing tests using a novel protocol for quantitative detection of water-soluble vitamin folic acid (FA, vitamin B9, vitamin M) as a prominent example of a low molecular weight compound (441.4 Da). Being a parent of a group of enzyme cofactors (referred to as folates), which play a significant role in the formation of purines, pyrimidines and methionine, FA is involved in DNA, RNA and protein biosynthesis, while the deviation of its level from normal values (3-20 ng/mL or 6.8 – 45.3 nM in human serum) can cause major health problems, including anaemia, psychiatric disorders, cardiovascular and cerebrovascular diseases, carcinogenesis or neuronal tube defects in new-borns. As the sample serum is typically limited (especially in newborns), the methods for FA diagnosis should be extremely sensitive, while the monitoring of FA level should be carried out in a relatively wide dynamic range. For these tests, we used a TDM shown in Fig. 3, which provides a fairly high sensitivity to bulk RI variations ($1.3 \cdot 10^4$ nm/RIU). The detection was implemented in a competitive assay mode which is beneficial for small analyte such as FA. A schematic of gold surface modification and a related experimental procedure are shown in Fig. 4a,b (more detailed experimental description is presented in Supplementary Information).

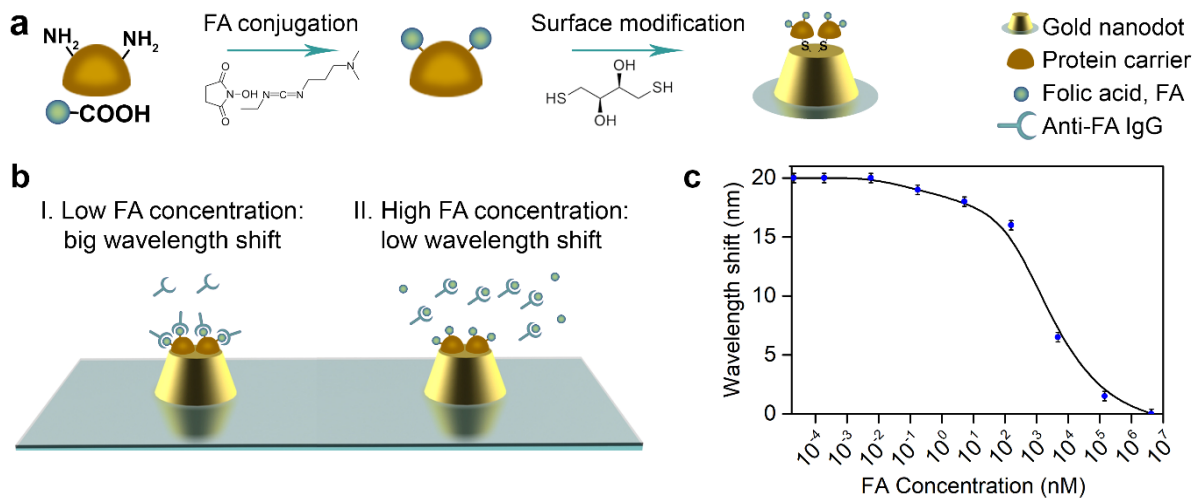

**Figure 4: Label-free detection of folic acid using TDM in conditions of the "scissor effect".** **a** Schematic illustration of gold surface modification with carrier protein conjugated to FA for the implementation of competitive FA detection. First, carrier protein (bovine serum albumin) is conjugated to FA via carbodiimide chemistry. Then, protein-FA is incubated with DTT to reduce –SH groups to enable gold surface biomodification. **b** Competitive label-free assay illustration for the detection of FA. A sample under investigation with FA is pre-incubated with anti-FA IgG. Next, the obtained complexes FA*anti-FA IgG are pumped through the liquid cell with gold nanodots coated with FA. When FA concentration is low (I), all FA binding sites on gold surface are coated with anti-FA antibodies thus resulting in sufficiently big detected signal. When the FA concentration is high enough (II), all anti-FA are pre-blocked with FA, so binding of IgG to the gold surface is impossible, thus resulting in small detecting signal. **c** Dependence of the spectral position of the minimum of reflection on a FA concentration in nM.

In the experiments, FA concentration in the solution pumped through liquid cell was varied between $2·10^{-5}$ nM and $4·10^6$ nM. As shown in Fig. 4c, the change of FA concentration led to a gradual spectral shift of the minimum of reflection associated with TD by about 20 nm and at concentration of 5 pM of FA come to a saturation. It is important that the sensing response was linear in a wide range of FA concentrations from 5 nm to 5000 nm (3-log range), which is crucial for tasks of FA monitoring. The limit of detection (LOD) in competitive assays is normally determined via $2·\sigma$ criterion: LOD = A $- 2·\sigma$, where A is the maximal signal on the saturation, $\sigma$ is the error in the measurement at the last "zero" point. Substituting the value of $\sigma$ for spectral measurements (0.4 nm), we can find that spectral LOD was equal to 0.125 nM. Such a level of LOD seems to drastically outperform reported LOD values for all optical biosensor counterparts. Indeed, this value is at least 20 times lower than reported for label-based (0.38 µM[26], 80 nM[27], 2.9 nM under nanoparticle-enhanced SPR[28]) and label-free (2.3 nM under SPR[29] and 100 nM under LPR[30]) sensors, as well as orders of magnitude lower than reported in most alternative approaches (electrochemical, capillary electrophoresis, etc.)[31]. It is interesting to note that the obtained 20-fold gain of sensitivity compared to SPR[29] could not be explained solely by the increase of bulk sensitivity to RI variations, as the studied sample exhibited only slightly higher sensitivity ($1.3 \cdot 10^4$ nm/RIU as compared to $(2-5) \cdot 10^3$ nm/RIU observed under SPR at analogous spectral range). We suppose that such a gain is due to stronger electric field probing target molecules, as well as to better localization of electromagnetic fields observed for the case of nanostructured TDM, as compared to flat gold surface under SPR. It should be noted that the recorded LOD was limited by the spectral sensitivity of TDM ($1.3 \cdot 10^4$ nm/RIU) conditioned by the efficiency of "scissor effect" for a concrete experimental layout. As we showed above, one can design TDMs providing sensitivities $10^5$-$10^6$ nm/RIU and higher, which are unimaginable within current optical transduction biosensing technology. It is worth noting that high sensitivity associated with topological darkness effects described here could already been observed in some previous studies, but these effects were not clearly identified and properly explained. In particular, an anomalously high sensitivity of $3.2 \cdot 10^4$ nm/RIU was reported using plasmonic nanorod metamaterial composed of a "forest" of long Au nanorods (400-450 nm) arranged perpendicularly on a glass substrate[32], while sensitivity of $2.4 \cdot 10^3$ nm/RIU was observed in 3D woodpile-based plasmonic crystal metamaterial[33]. A detailed analysis of the experimental data reported in these works shows that in both cases the quasi-effective media provided almost zero intensity in reflection under a relatively narrow range of angle variations, which is the telltale feature of TD phenomenon.

**New paradigm in label-free biosensing**

As the main message of the presented study, by implementing the "scissor effect" via a proper design of topologically dark metamaterials, we theoretically propose a new paradigm in label-free optical biosensing, which promises virtually unlimited sensitivity. Indeed, conventional paradigm in such biosensing modality implies the use of resonant phenomena (SPR, LPR, etc.) and following the resonant curve with respect to its initial position due to biological binding events on the sensor surface. Instead, we propose to use the phenomenon of topological zero reflection, realized with the help of a TDM mimicking effective media with appropriate optical constants $n$ and $k$, and study the spectral shifts of this zero reflection position with respect to the spectral curve $n(\lambda)$ and $k(\lambda)$ of the TDM. We showed that under a proper design of TDM, the "scissor effect" conditioned by a very small angle of intersection of the spectral curve and zero reflection surface can be realized, which leads to orders of magnitude increase of biosensor sensitivity. We also described a pathway on how to design UTDM which relies on the possibility to design a metamaterial with given spectral properties. It is important that UTDMs can be realized using not only metal, but also semiconductor or dielectric structures, while local structure imperfections appear to be not significant due to topologically protected zero reflection. Therefore, in contrast to many other metamaterial structures, they can be fabricated by cost-efficient methods such as nanoparticle lithography or self-assembling. Our work opens endless opportunities for theoreticians to design hypersensitive TDMs for biosensing (and other applications) as well as novel exciting possibilities for experimentalists to realize these UTDMs using nanostructured and heterostructured (meta)materials.

In the view of potential applications, we demonstrated the possibility of FA monitoring using a TDM, which provided a record sensitivity combined with a linear response in a wide dynamic range. TDMs can be easily adapted for more intelligent fundamental and biomedical applications or for highly sensitive detection of other low molecular weight compounds and proteins, including clinically relevant antibiotics, toxins, hormones, disease-related antibodies. It is also important that UTDMs can be properly structured to offer sensor surface for the immobilization of multiple receptors, which opens up avenues for high throughput analyses of numerous lead analytes in parallel. For example, ultrasensitive detection of low-molecular weight compounds (such as various carbohydrate metabolites, homocysteine, cholesterol, vitamins, various hormones) is necessary in clinical practice for analyzing the current stage and monitoring the dynamics of the disease in patients. The development of reliable and sensitive methods for analyzing illegal drugs or doping in sports is also of high importance. As for the food industry, different methods of analysis with significant sample dilution (designed to eliminate the matrix effect) and without sophisticated sample preparation for the detection of pathogens, antibiotics or vitamins, are also extremely popular. Taking into account the relatively low cost of both UTDMs and required hardware, the UTDM-based biosensing technologies represent an appealing platform for the new-generation express point-of-care testing, for, e.g., COVID-19 rapid analysis or intraoperative diagnostics.

To conclude, we report the "scissor effect" using a new type of metamaterials, which can provide extremely high (theoretically unlimited) spectral sensitivity in the detection of biological binding events and thus allows one to solve bottleneck sensitivity limitation problem of current optical label-free biosensing technology. We provided an algorithm for designing ultrasensitive TDMs with any given spectral sensitivity. Using the "scissor effect", we experimentally demonstrated the detection of folic acid in the wide 3-log linear dynamic range with the limit of detection of 0.125 nM, which is orders of magnitude better than previously reported for all optical counterparts. Our work paves way to robust, inexpensive, fast and accessible label-free optical biosensors.

## Materials and Methods

### Device fabrication

High-quality regular and homogenous arrays of gold coupled dot pairs were produced by e-beam lithography on a clean microscopic glass substrate covered by a thin Cr (5nm) sublayer (routinely used to avoid charging during electron beam lithography). We employed a double-layered resist (80 nm of 3% 495 polymethyl methacrylate (PMMA) for the bottom resist layer and 50 nm of 2% 950 PMMA for the top layer) in order to improve the subsequent lift-off process. The exposure was performed using a LEO-RAITH e-beam lithography system followed by development in 1:3 methyl isobutyl ketone (MIBK):isopropanol (IPA) developer for 30 s. After lithography, we deposited 5nm of Cr (to improve adhesion) and 90nm Au by electron beam evaporation with the help of a Moorfield system. Our deposition rate was controlled precisely at 1.0 ˚A s$^{-1}$ and the base pressure was 1.0 × 10$^{-6}$ Torr. The thickness of growing metal film was monitored by a calibrated quartz microbalance (CQM). For the lift-off procedure, the sample was immersed in acetone for approximately 1 h. Finally, a scanning electron microscopy (SEM) image of the fabricated double nanodots structure was taken to determine the size of dots, periodicity of nanostructure and separation between dots in the pair. The fabrication of our samples is described in more detail in our previous works[11].

### Ellipsometric parameters $\Psi$ and $\Delta$

Ellipsometry is a sensitive method that can be used to measure optical properties of materials. Ellipsometry routinely provides the amplitude ($\Psi$) and the phase ($\Delta$) parameters for light reflected from an object. These parameters are related to the complex reflected field amplitudes $r_p = \frac{E_p}{E_i}$ and $r_s = \frac{E_s}{E_i}$ (where $E_i$ is the incident light, $E_p$ and $E_s$ are the reflected fields for *p*- and *s*- polarizations, respectively) by the following equation $\rho = \frac{r_p}{r_s} = \tan(\Psi)\exp(i\Delta)$ [34]. The function $\Psi$ represents the modulus of the ratio of Fresnel reflection amplitudes for *p*- and *s*-polarizations, while $\Delta$ provides the phase shift between *p*-and *s*-components of the light. A spectroscopic ellipsometer can measure the dependence of $\Psi$ and $\Delta$ on light wavelength. In addition, a variable angle ellipsometer allows one to measure the spectral dependences of $\Psi$ and $\Delta$ on angle of incidence. Intensity reflections and transmissions ($R_p$, $T_p$) for *p*- and ($R_s$, $T_s$) for *s*-polarized light at various angles of incidence can also be measured. The measurements of TDM nanostructures were performed across a wavelength range of 240-1700nm with the help of a variable angle spectroscopic ellipsometer (VASE) M-2000F, manufactured by J.A. Woollam, using a rotating compensator-analyzer configuration.

## Acknowledgements

A.N.G. and V.G.K. acknowledge support of Graphene Flagship programme, Core 3 (881603).

## Author contributions

A.N.G. conceived the research and made theoretical simulations of UTDM structures. V.G.K., G.T., A.D, V.O.S., S.M.D., A.V.K, A.N.G. designed and performed experiments. All authors analysed and discussed obtained data. A.N.G. and A.V.K. prepared the manuscript using data from co-authors. A.N.G. and A.V.K. guided the project. All authors have given approval to the final version of the manuscript.

## Additional Information

Supplementary Information is available. Correspondence and request for materials should be addressed to A.N.G.

## Competing financial interests

The authors declare no competing financial interests.

## References

1   Banica, F.-G. *Chemical Sensors and Biosensors: Fundamentals and Applications* (John Wiley & Sons, 2012).
2   Liedberg, B., Nylander, C. & Lundström, I. Biosensing with surface plasmon resonance — how it all started. *Biosensors and Bioelectronics* **10**, i-ix, doi:https://doi.org/10.1016/0956-5663(95)96965-2 (1995).
3   Homola, J. Surface Plasmon Resonance Sensors for Detection of Chemical and Biological Species. *Chemical Reviews* **108**, 462-493, doi:10.1021/cr068107d (2008).
4   Anker, J. N. *et al.* Biosensing with plasmonic nanosensors. *Nature Materials* **7**, 442-453, doi:10.1038/nmat2162 (2008).
5   Maier, S. A. *Plasmonics: fundamentals and applications*. (Springer Science & Business Media, 2007).
6   Wang, F. & Shen, Y. R. General properties of local plasmons in metal nanostructures. *Physical review letters* **97**, 206806 (2006).
7   Kravets, V., Schedin, F. & Grigorenko, A. Extremely narrow plasmon resonances based on diffraction coupling of localized plasmons in arrays of metallic nanoparticles. *Physical review letters* **101**, 087403 (2008).
8   Kravets, V. G., Kabashin, A. V., Barnes, W. L. & Grigorenko, A. N. Plasmonic Surface Lattice Resonances: A Review of Properties and Applications. *Chemical Reviews* **118**, 5912-5951, doi:10.1021/acs.chemrev.8b00243 (2018).
9   Kabashin, A. V. & Nikitin, P. I. Interferometer based on a surface-plasmon resonance for sensor applications. *Quantum Electronics* **27**, 653-654, doi:10.1070/qe1997v027n07abeh001013 (1997).
10  Kabashin, A. V., Patskovsky, S. & Grigorenko, A. N. Phase and amplitude sensitivities in surface plasmon resonance bio and chemical sensing. *Opt. Express* **17**, 21191-21204, doi:10.1364/OE.17.021191 (2009).
11  Kravets, V. G. *et al.* Singular phase nano-optics in plasmonic metamaterials for label-free single-molecule detection. *Nature Materials* **12**, 304-309, doi:10.1038/nmat3537 (2013).
12  Malassis, L. *et al.* Topological Darkness in Self-Assembled Plasmonic Metamaterials. *Advanced Materials* **26**, 324-330, doi:https://doi.org/10.1002/adma.201303426 (2014).
13  Wu, F. *et al.* Layered material platform for surface plasmon resonance biosensing. *Scientific reports* **9**, 1-10 (2019).
14  Kravets, V. G., Schedin, F., Kabashin, A. V. & Grigorenko, A. N. Sensitivity of collective plasmon modes of gold nanoresonators to local environment. *Opt. Lett.* **35**, 956-958, doi:10.1364/OL.35.000956 (2010).
15  Danilov, A. *et al.* Ultra-narrow surface lattice resonances in plasmonic metamaterial arrays for biosensing applications. *Biosensors and Bioelectronics* **104**, 102-112 (2018).
16  Kabashin, A. V. *et al.* Phase-Responsive Fourier Nanotransducers for Probing 2D Materials and Functional Interfaces. *Advanced Functional Materials* **29**, 1902692, doi:https://doi.org/10.1002/adfm.201902692 (2019).
17  Choy, T. C. *Effective medium theory: principles and applications*. Vol. 165 (Oxford University Press, 2015).

18   Kravets, V. G., Schedin, F. & Grigorenko, A. N. Plasmonic blackbody: Almost complete absorption of light in nanostructured metallic coatings. *Physical Review B* **78**, 205405, doi:205405

10.1103/PhysRevB.78.205405 (2008).

19   Liu, N., Mesch, M., Weiss, T., Hentschel, M. & Giessen, H. Infrared Perfect Absorber and Its Application As Plasmonic Sensor. *Nano Letters* **10**, 2342-2348, doi:10.1021/nl9041033 (2010).

20   Leonhardt, U. Optical Conformal Mapping. *Science* **312**, 1777-1780, doi:10.1126/science.1126493 (2006).

21   Pendry, J. B., Schurig, D. & Smith, D. R. Controlling Electromagnetic Fields. *Science* **312**, 1780-1782, doi:10.1126/science.1125907 (2006).

22   Leonhardt, U. & Philbin, T. G. in *Progress in Optics* Vol. 53 (ed E. Wolf) 69-152 (Elsevier, 2009).

23   Chen, H., Chan, C. T. & Sheng, P. Transformation optics and metamaterials. *Nature Materials* **9**, 387-396, doi:10.1038/nmat2743 (2010).

24   Huidobro, P. A., Nesterov, M. L., Martín-Moreno, L. & García-Vidal, F. J. Transformation Optics for Plasmonics. *Nano Letters* **10**, 1985-1990, doi:10.1021/nl100800c (2010).

25   Pendry, J. B., Aubry, A., Smith, D. R. & Maier, S. A. Transformation Optics and Subwavelength Control of Light. *Science* **337**, 549-552, doi:10.1126/science.1220600 (2012).

26   Zhang, W. *et al.* Carbon quantum dots as fluorescence sensors for label-free detection of folic acid in biological samples. *Spectrochimica Acta Part A: Molecular and Biomolecular Spectroscopy* **229**, 117931 (2020).

27   Liu, S., Hu, J. & Su, X. Detection of ascorbic acid and folic acid based on water-soluble CuInS 2 quantum dots. *Analyst* **137**, 4598-4604 (2012).

28   Cao, Y., Griffith, B., Bhomkar, P., Wishart, D. S. & McDermott, M. T. Functionalized gold nanoparticle-enhanced competitive assay for sensitive small-molecule metabolite detection using surface plasmon resonance. *Analyst* **143**, 289-296 (2018).

29   Indyk, H. E. *et al.* Determination of biotin and folate in infant formula and milk by optical biosensor-based immunoassay. *Journal of AOAC International* **83**, 1141-1148 (2000).

30   Ahmad, R. *et al.* Water-soluble plasmonic nanosensors with synthetic receptors for label-free detection of folic acid. *Chemical Communications* **51**, 9678-9681, doi:10.1039/C5CC01489A (2015).

31   Batra, B., Narwal, V., Kalra, V., Sharma, M. & Rana, J. Folic acid biosensors: A review. *Process Biochemistry* **92**, 343-354 (2020).

32   Kabashin, A. V. *et al.* Plasmonic nanorod metamaterials for biosensing. *Nature materials* **8**, 867-871 (2009).

33   Aristov, A. I. *et al.* 3D plasmonic crystal metamaterials for ultra-sensitive biosensing. *Scientific reports* **6**, 1-8 (2016).

34   Azzam, R. M. A. & Bashara, N. M. *Ellipsometry and Polarized Light*. (North-Holland, 1977).